# Negative Thermal Expansion in cubic ZrW$_2$O$_8$: Role of Phonons in Entire Brillouin Zone From ab-inito Calculations

M. K. Gupta, R. Mittal and S. L. Chaplot
*Solid State Physics Division, Bhabha Atomic Research Centre, Trombay, Mumbai 400 085, India*

We report ab-inito density functional theory calculation of phonons in cubic phase of ZrW$_2$O$_8$ in the entire Brillouin zone and identify specific anharmonic phonons that are responsible for large negative thermal expansion (NTE) in terms of translation, rotation and distortion of WO$_4$ and ZrO$_6$. We have used density functional calculations to interpret the experimental phonon spectra as a function of pressure and temperature as reported in literature. We discover that the phonons showing anharmonicty with temperature are different from those showing anharmonicity with pressure although both are of similar frequencies. Only the latter phonons are associated with NTE. Therefore the cubic and/or quadratic anharmonicity of phonons is not relevant to NTE but just the volume dependence of frequencies. The calculations are able to reproduce the observed anomalous trends, namely, the softening of the low frequency peak at about 5 meV in the phonon spectra with pressure and its hardening with temperature, while both the changes involve a compression of the lattice.

The discovery of large isotropic negative thermal expansion (NTE) in cubic phase of ZrW$_2$O$_8$ two decades ago has lead to great excitement in the field of material science. Since then anamolous thermal expansion behavior has been found in large number of open frame work compounds [1-3]. The thermal expansion is found to show large variation with increasing framework flexibility. ZrW$_2$O$_8$ has M-O-M' (M, M'= Zr, W) type of linkages and shows [1] negative volume thermal expansion coefficient of -29 × 10$^{-6}$ K$^{-1}$ at 300 K. Increasing flexibility in the structure has lead to the discovery of compounds exhibiting [4] colossal positive and negative thermal expansion. The compounds find applications in forming the composites with tailored thermal expansion coefficients useful for applications such as in fiber optic communication systems.

At ambient temperature ZrW$_2$O$_8$ crystallizes [1] in cubic structure (P2$_1$3, Z=4) that consists of ZrO$_6$ octahedral and WO$_4$ tetrahedral units. Diffraction, spectroscopic as well as computer simulations techniques [5-12] have been used to understand the thermodynamic behaviour of ZrW$_2$O$_8$. All these works show that anharmonicity of low energy phonon modes has major contribution to the observed thermal expansion behavior. X-ray absorption fine structure (XAFS) technique has been used to investigate the local structure of ZrW$_2$O$_8$ [12] as a function of temperature. The XAFS measurements led to a suggestion that NTE in ZrW$_2$O$_8$ could be due to the translational motion of WO$_4$ tetrahedra in a plane perpendicular to the [111] axis along with the correlated motion of three nearest ZrO$_6$ octahedra. The reverse Monte Carlo analysis of the neutron total scattering data suggested [10] that WO$_4$ as well as ZrO$_6$ polyhedra rotate and translate as rigid units. The reverse Monte Carlo analysis does not support the interpretations made from the XAFS data. Hancock et al [11] proposed other modes involving translation and rotation of polyhedra. It seems all the phonon modes identified from various techniques could contribute to NTE.

Earlier neutron scattering data [9] as well as theoretical [8] estimates of the anharmonicty of the phonons in ZrW$_2$O$_8$ using interatomic potential model indicated that modes of energy below 8 meV are responsible for observed NTE. However estimates based on the Raman spectroscopy showed [7] that several modes nearly upto 50 meV contribute to NTE. The large disagreement in the energy range as well as nature of the low-energy modes in previous works indicated the need for understanding of NTE behaviour in ZrW$_2$O$_8$ using ab-inito calculations. Recently ab-initio calculations of zone center phonon modes have been published [6]. However the authors concluded [6] that one should fully explore the nature of the phonons in the entire Brillouin zone for understanding the mechanism of NTE in cubic ZrW$_2$O$_8$. Here we report such a comprehensive calculations and identify specific zone-boundary modes that are highly anharmonic. The calculations are able to reproduce the observed NTE in cubic ZrW$_2$O$_8$ as well as anomalous trends of the phonon spectra with increase in temperature and pressure.

Important soft modes were identified in cubic ReO$_3$ [13] and ScF$_3$ [14] at M and R-points in the Brillouin zone respectively. These modes show simultaneously both large negative Grüneisen parameter as well as large quadratic anharmonicity, the former leading to NTE and latter to the temperature dependence. In case of ZrW$_2$O$_8$, we find that the modes that show large negative Grüneisen parameter contribute to NTE are not necessarily the same as those showing cubic and/or quadratic anharmonicity and significant temperature dependence. This finding means that the modes found anharmonic in temperature dependent measurements are not necessarily relevant to NTE.

The first-principle calculations of lattice dynamics have been performed using Vienna Ab-initio Simulation Package (VASP) [15,16]. The generalized gradient approximation (GGA) exchange correlation given by Perdew Becke and Ernzerhof [17,18] with projected–augmented wave method has been used. The phonon spectra have been calculated using PHONON5.2 software developed by Parlinski [19]. The details about the ab-initio calculations are given in supplementary material [20]. The ab-initio calculations reproduce the equilibrium crystal structural parameters and mean squared amplitude of various atoms quite satisfactorily as given in Table SI of the supplementary material [20].

The calculated phonon spectrum is found to be in excellent agreement with the experimental phonon spectrum [5] as shown in Fig. 1. The calculated energies of all the zone centre modes are also shown in Fig. 1. The calculated partial density of states of various atoms shows (Fig. S1 of [20]) that vibrations due to Zr atoms span only up to 50 meV, while vibrations due to W and O span the entire energy range. The W-O stretching modes lie in the energy range from 85 -130 meV.

The calculated phonon dispersion relation along the high symmetry directions is shown in Fig. S2 [20]. The low energy range of phonon dispersion up to 50 meV contains large number of non-dispersive phonon branches, which give rise to several peaks in density of states. To emphasis the anharmonic nature of low energy phonons, we have also shown the phonon dispersion up to 10 meV (Fig. 2) at 0 and 1 kbar. We find that several phonon branches soften with increasing pressure. The lowest optic mode is calculated at 40 cm$^{-1}$ (~5 meV), which is in excellent agreement with the experimental value of 40 cm$^{-1}$ from Raman [7] as well as infra-red measurements [11]. The optic modes along with several phonon branches give rise to the first peak in the calculated phonon density of states at 4.5 meV which is observed at 3.8 meV in neutron scattering experiments [5]. The low-energy peak also leads to a sharp increase in the specific heat at low temperatures (Fig. S3 Ref. [20]). The good agreement between our calculated and experimental specific heat [21, 22] supports the correctness of the low energy phonon density of states provided by the ab-initio calculations.

The calculation of thermal expansion is carried out using the quasi-harmonic approximation. Each phonon mode of energy $E_i$ contributes to the volume thermal expansion coefficient [23] that is given by the relation $\alpha_V = \frac{1}{BV} \sum_i \Gamma_i C_{Vi}(T)$, where $V$ is the unit cell volume, $B$ is the bulk modulus, $\Gamma_i$ (=-$\partial \ln E_i/\partial \ln V$) are the mode Grüneisen parameters and $C_{Vi}$ the specific-heat contributions of the phonons of energy $E_i$. The index $i$ run over the various phonon branches and all the wavevectors in the Brillouin zone. The Grüneisen parameter $\Gamma_i$, are numerically calculated from the pressure dependence of phonon modes around ambient pressure and 1 kbar.

The calculated $\alpha_V$ at 300 K from ab-initio calculation is -22.5 × 10$^{-6}$ K$^{-1}$, while the experimental value [1] is about -29 × 10$^{-6}$ K$^{-1}$. The calculated relative volume thermal expansion is shown in Fig. 3(a). The discontinuity in the experimental data at about 400 K is associated with a disorder phase transition. We find that there is a slight deviation between the experimental data [1] and the calculations at low temperatures due to the underestimation of the contribution from low energy phonon modes. Similar underestimation of the anharmonicity of low energy phonon modes is also found in cases of Zn(CN)$_2$ [24] as well as Ag$_3$M(CN)$_6$ (M=Co,Fe) [25]. The properties of the low energy phonon modes are very sensitive to volume of the crystal. DFT calculations overestimate or underestimate crystal volume depending on the exchange correlation functional. The elastic constants values as calculated from the slopes of the acoustic modes along [100] and [110] are $C_{11}$=137.6 GPa, $C_{44}$= 24.9 GPa and $C_{12}$= 66.7 GPa, which are in fair agreement with the experimental data [26] of 161.8 GPa, 29.4 GPa and 75.5 GPa respectively.

The contribution of phonon density of states at energy E to the thermal expansion has been determined (Fig. 3(b)) as a function of phonon energy at 300 K. We find that the maximum contribution to $\alpha_V$ is found to be from phonon modes of energy 4.5 ± 1 meV which is consistent with the previous analysis of high pressure inelastic neutron scattering measurements [9] as well as diffraction data [5].

The eigenvectors of a few of the low energy modes (Table I) that contribute most to NTE have also been plotted (Fig. 4 and S4). The nature of these phonons can be best understood by the animations which are available in the supplementary material [20]. The lowest Γ-point mode of 4.93 meV ($\Gamma_i$ = -7.0) involves out-of-phase translation of two chains consisting of WO$_4$ and ZrO$_6$, while the Γ-point mode of 5.21 meV ($\Gamma_i$ = -5.7) show out-of-phase rotation of WO$_4$ and translation of ZrO$_6$ in two different chains. These modes also involve significant distortion of WO$_4$ tetrahedra formed around W1 and W2.

Hancock et al [11] proposed two types of modes for understanding the mechanism of NTE. In one of the mode both ZrO$_6$ as well as WO$_4$ in a chain rotate and also translate along the <111> axis. As discussed above, for the two lowest optic modes we have not found simultaneous rotational motion of both the ZrO$_6$ as well as WO$_4$. However we find that for the X-point modes of 3.90 meV ($\Gamma_i$ = -5.7) and 4.16 meV ($\Gamma_i$ = -2.4), the motion of polyhedral units is similar to that proposed by Hancock et al [11]. The modes show in-phase translation and rotation of WO$_4$ and ZrO$_6$ in a single chain. The motion of tetrahedral and octahedral units in two different chains is also in-phase. While the two modes seem to be of similar nature, the relative amplitudes of Zr, W atoms and O atoms are found to be different.

The second mode proposed by Hancock et al [11] indicates that ZrO$_6$ octahedron rotates opposite to the WO$_4$ tetrahedra. We find that R-point (0.5 0.5 0.5) mode of 5.29 meV with $\Gamma_i$ value of -11.7 show similar behaviour. The two WO$_4$ around W1 and W2 in a chain rotate in-phase while ZrO$_6$ rotate out-of phase. In general we find that in most of the modes, amplitude of the free oxygens O3 and O4 are larger as compared to that of shared oxygens O1 and O2. This means that rotation of WO$_4$ and ZrO$_6$ is accompanied by distortion of these polyhedra.

The M-point modes of 4.51 meV and 4.65 meV energy have negative Grüneisen parameter $\Gamma_i$ value of about -12.7 and -12.8 respectively. The mode at 4.51 meV involves in-phase translation and bending of WO$_4$ and ZrO$_6$ network. The mode is very similar to that previously described by Cao et al [12], where a correlated motion between WO$_4$ and it nearest ZrO$_6$ is shown to lead NTE. However, for 4.65 meV mode we find out-of-phase translation of WO$_4$ and ZrO$_6$ in two chains.

The temperature dependence of phonon density of states of cubic $ZrW_2O_8$ shows [5] hardening of the peak at 3.8 meV on increase of temperature from 50 K to 300 K. On the other hand, the same peak is found to soften with pressure [9] although both increase in pressure and temperature involve compression of the lattice. Temperature and pressure variation of the phonon energy is known to occur due to anharmonicity of the interatomic potential. The change in phonon energies is due to two effects. The so called "implicit" anharmonicity, refers to the volume dependence of the phonon spectra that can be calculated in the quasiharmonic approximation. The second is the "explicit" anharmonicity, which refer to the changes in phonon frequencies due to large thermal amplitude of atoms. The change in phonon energies with temperature is due to both the "implicit" as well as "explicit" anharmonicities, while the pressure effect only involves the implicit part. We would also call the "implicit" and "explicit" parts as volume and amplitude effects respectively.

In a complex crystal it is quite difficult to estimate the anharmonic effects rigorously. However, one can make certain simplifying assumptions and arrive at qualitative trends in the shifts of selected phonons as a function of temperature. The potential wells of a few of the phonon modes at high symmetry points in the cubic Brillouin zone have been calculated and used to estimate the temperature dependence of the phonon frequencies. The detailed procedure for calculation of explicit part of the temperature dependence of phonon modes can be found in supplementary material [20] as well as in Refs. [27-29].

The potential well (Fig.5 and S5) of the seven modes of energy around 5 meV, along the high symmetry points namely $\Gamma$, X, M and R in the cubic Brillouin zone, have been calculated. The energy of modes may increase or decrease with increase of temperature, depending on the nature of anharmonicity. The potential wells for $\Gamma$ point mode of energy at 4.93 and 5.21 meV (TABLE SII) have cubic as well as quadratic anharmonicity, while all the remaining five modes have only quadratic anharmonicity. The potential well for M point mode of 4.65 meV with Grüneisen parameter $\Gamma_i$ value of about -12.7 has also been plotted at 1 kbar. As expected the width of the well is slightly increased due to softening of phonon mode on compression.

The anhamonicity parameters (Table SII) as obtained from fitting of equation (S1) to the potential well are used for calculating the temperature dependence of phonon modes. We find (Fig. 6 and S6) that zone boundary mode of energy 4.16 meV (0 K) at X-point shows maximum hardening and shifts to 4.78 meV on increase of temperature to 300 K. The low energy $\Gamma$-point modes do not respond to temperature and remain nearly invariant with temperature. The R-point mode of energy 5.29 meV shows normal behaviour of decrease of phonon energy with increase of temperature. The calculated energy shift for low energy modes on increase of temperature from 0 to 300 K is given in Table I. Ab-initio calculations are able to qualitatively explain the experimentally observed [5] temperature dependence of low energy phonon spectra of cubic $ZrW_2O_8$, which indicates that a peak in the phonon density of states at 3.8 meV shifts to 4.05 meV on increase of temperature from 50 K to 300 K.

We would like to draw attention to the fact that the modes at M and R point show large implicit anharmonicity. These modes are important for understanding the NTE behaviour. However as far as temperature dependence is concerned, the X-point mode having low negative Grüneisen parameter $\Gamma_i$ value of -2.4 shows maximum temperature dependence. Recently in case of NTE compounds $ScF_3$ [14] and $ReO_3$ [13] respectively, R-point and M-point modes are found to show large pressure as well as temperature dependence. The authors also found large quadratic anharmonicity for the same modes. We would like to say that quadratic anharmonicity is useful to explain the large temperature dependence of R-point and M-point modes and is not relevant to NTE.

To summarize, the ab-initio density functional calculations of phonons modes of $ZrW_2O_8$ have been reported in the entire Brillouin zone. Certain phonon modes are found to be highly anharmonic in nature. The thermal expansion behaviour is calculated from the volume dependence of phonon energies. The calculations agree quite well with the reported NTE behavior of $ZrW_2O_8$. We have also been able to explain the observed anomalous pressure as well as temperature variation of the energies of phonon modes. The increase of the frequency with temperature essentially results from the cubic and/or quadratic anharmonic part of the phonon potential, which is able to explain the temperature dependence of low energy modes as reported in the literature.

TABLE I The calculated change in energy of low energy phonon modes on increase of temperature from 0 to 300 K. Γ, X, M and R are the high symmetry points in the cubic Brillouin zone. $E_i$ and $\Gamma_i$ are the energy and Grüneisen parameter of $i^{th}$ mode mode, respectively at 0 K. While $\Delta E_V$, $\Delta E_A$ and $\Delta E_T$ are the change in energy of mode due to change in volume (implicit anharmonicity), increase in thermal amplitudes of atoms (explicit anharmonicity) and total change in energy of mode on increase of temperature from 0 to 300 K. The parameters E, $\Delta E_V$, $\Delta E_A$ and $\Delta E_T$ are in "meV" units.

| Representation | $E_i$ | $\Gamma_i$ | $\Delta E_V$ | $\Delta E_A$ | $\Delta E_T$ |
|---|---|---|---|---|---|
| Γ | 4.93 | -7.0 | -0.22 | 0.15 | -0.07 |
| Γ | 5.21 | -5.7 | -0.19 | 0.16 | -0.03 |
| X | 3.90 | -5.7 | -0.14 | 0.22 | 0.08 |
| X | 4.16 | -2.4 | -0.06 | 0.68 | 0.62 |
| M | 4.51 | -12.7 | -0.36 | 0.42 | 0.06 |
| M | 4.65 | -12.8 | -0.37 | 0.55 | 0.18 |
| R | 5.29 | -11.7 | -0.39 | -0.38 | -0.77 |

**FIG. 1** The calculated (0 K) and experimental (300 K) [5] neutron-weighted phonon spectra in the cubic phase of $ZrW_2O_8$. To account for the experimental resolution broadening in cubic $ZrW_2O_8$, the calculated spectra have been convoluted with a Gaussian of FWHM of 5 % and 8 % of the energy transfer below and above 40 meV, respectively. For better visibility the experimental phonon spectra is shifted along the y-axis by 0.03 meV$^{-1}$. The symbols represent the positions of the calculated zone centre optic modes. The A, E, F(TO) and F(LO) correspond to the group theoretical decomposition of the phonon modes at zone centre.

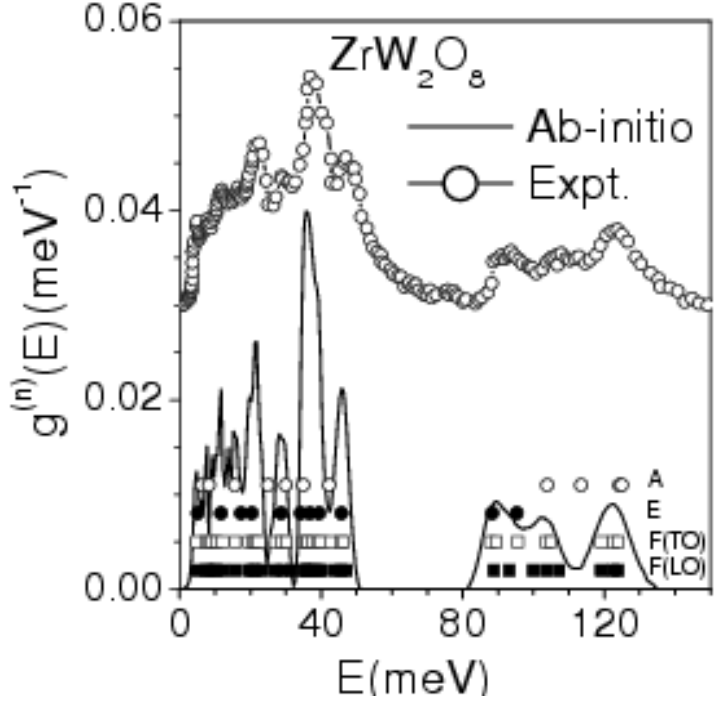

**FIG. 2** (Color online) The Calculated low-energy part of the pressure dependence dispersion relation for $ZrW_2O_8$. The solid and dashed lines correspond to the calculations at ambient pressure and 1 kbar. The Bradley-Cracknell notation is used for the high-symmetry points along which the dispersion relations are obtained. $\Gamma=(0,0,0)$; $X=(1/2,0,0)$; $M=(1/2,1/2,0)$ and $R=(1/2,1/2,1/2)$.

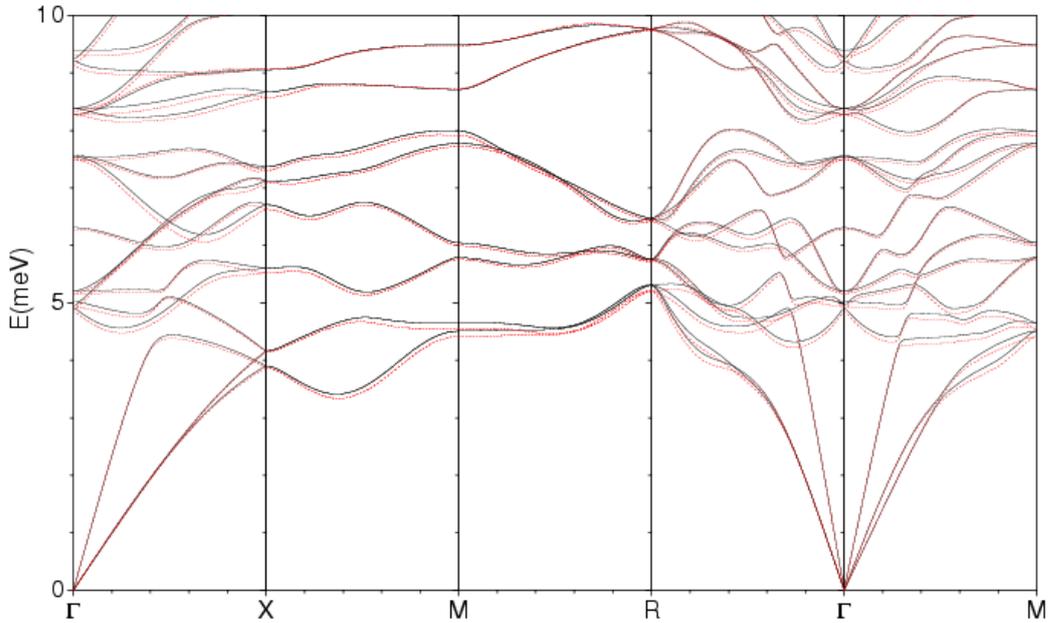

**FIG. 3 (a)** The calculated and experimental [1] relative volume thermal expansion for cubic $ZrW_2O_8$, $(V_T/V_{300}-1) \times 100\%$, $V_T$ and $V_{300}$ being the cell volumes at temperature T and 300 K respectively. There is a small sharp drop in volume for cubic phase at about 400 K associated with a order disorder phase transition. (b) The contribution of phonons of energy E to the volume thermal expansion as a function of E at 300 K from ab-initio as well as phonon data [9].

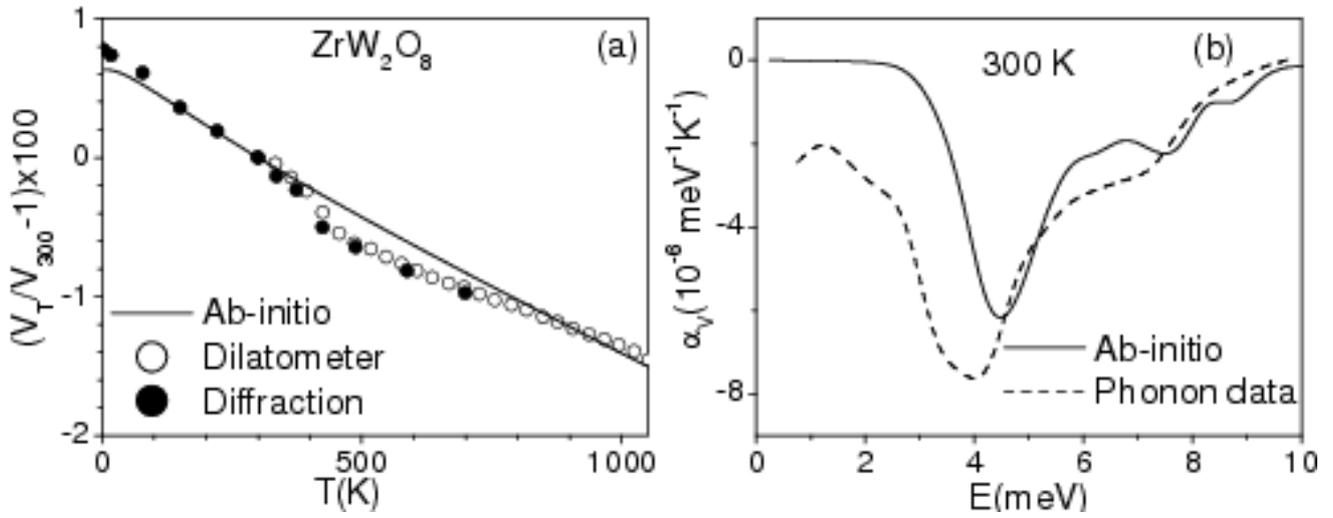

**Fig. 4** (Color online) Polarization vectors of low energy phonon modes in cubic $ZrW_2O_8$. Γ, X, M and R are the high symmetry points in the cubic Brillouin zone. The numbers after the mode assignments give the phonon energies and Grüneisen parameters respectively. The lengths of arrows are related to the displacements of the atoms. The atoms are labeled as indicated in Ref. [1].

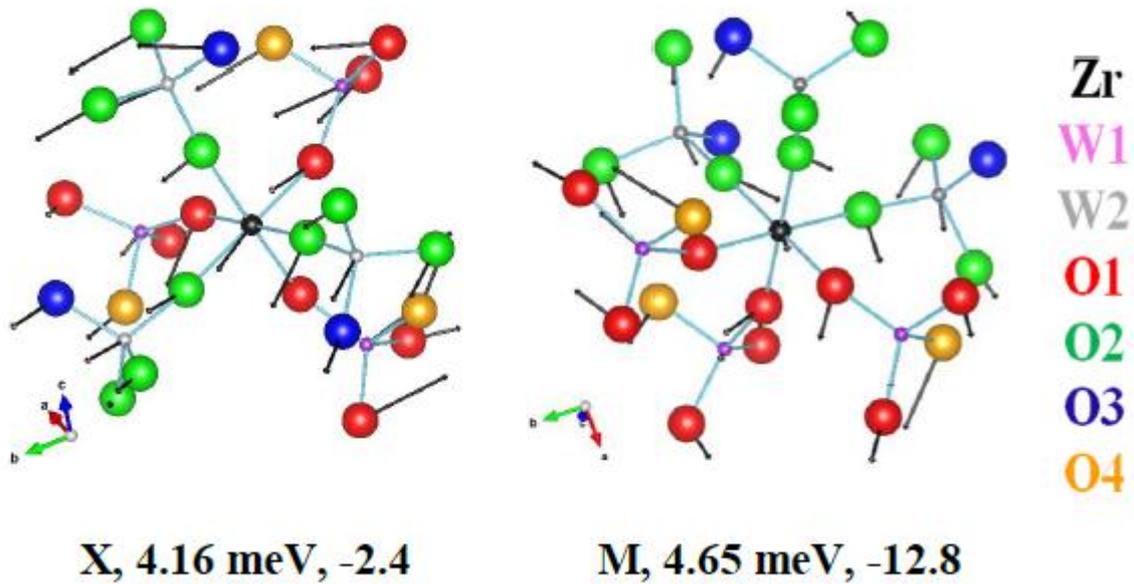

**Fig. 5** (Color online) The calculated potential wells of low energy phonon modes in cubic $ZrW_2O_8$. X and M are the high symmetry points in the cubic Brillouin zone. The numbers after the mode assignments give the phonon energies and Grüneisen parameters respectively.

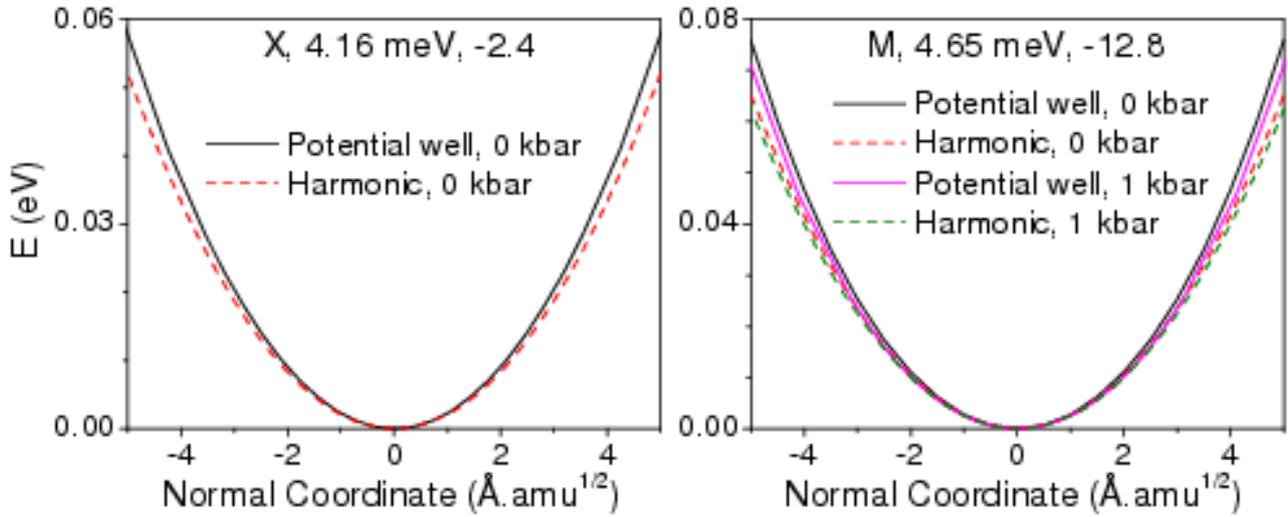

**Fig. 6** (Color online) The calculated temperature dependence of low energy phonon modes in cubic $ZrW_2O_8$. The temperature dependence of phonon modes is due to both the "implicit" as well as "explicit" anharmonicities. X and M are the high symmetry points in the cubic Brillouin zone. The numbers after the mode assignments give the phonon energies and Grüneisen parameters respectively. Solid circles are the data taken from the experimental temperature dependence of phonon density of states [5], which represents average over entire Brillouin zone. The line through the circles is guide to the eye.

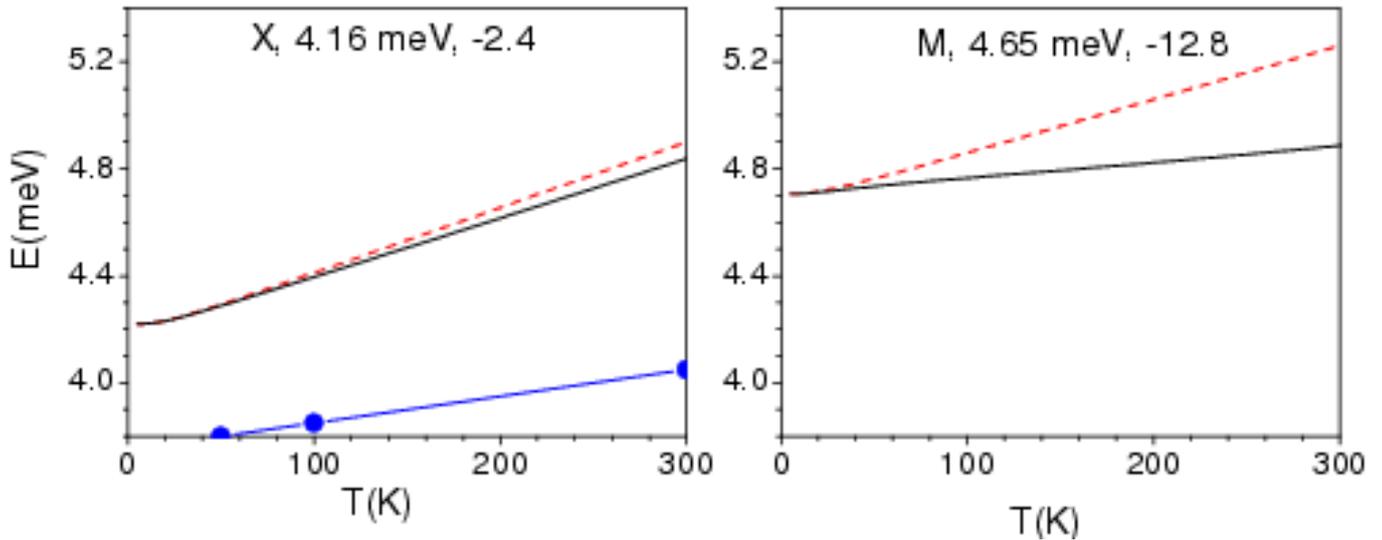

## Supplementary Material

## Negative Thermal Expansion in cubic ZrW$_2$O$_8$: Role of Phonons in Entire Brillouin Zone From ab-inito Calculations

M. K. Gupta, R. Mittal and S. L. Chaplot
*Solid State Physics Division, Bhabha Atomic Research Centre, Trombay, Mumbai 400 085, India*

**Details of ab-initio Calculations**

The first-principle calculations of lattice dynamics have been performed using Vienna Ab-initio Simulation Package (VASP) [1,2]. An energy cutoff of 820 eV has used for the plane wave basis, and integrations over the Brillouin zone of the cubic crystal were sampled with a 4 × 4 × 4 mesh of k-points generated by Monkhorst scheme [3].The energy cutoff and K mesh is found to be sufficient for total energy convergence of the order of meV. The generalized gradient approximation (GGA) exchange correlation given by Perdew Becke and Ernzerhof [4,5] with projected–augmented wave method has been used. A 2×2×2 super cell was found to be sufficient for phonon frequency convergence in entire Brillouin zone. Total energies and inter-atomic forces were calculated for the 22 structures resulting from individual displacements of the seven symmetry in equivalent atoms along the three cartesian directions (±x, ±y and ±z). The convergence criteria for the total energy and ionic forces were set to $10^{-8}$ eV and $10^{-5}$ eV Å$^{-1}$, respectively. The phonon spectra have been calculated using PHONON5.2 software developed by Parlinski [6].

**The calculation of Explicit part of the Temperature Dependence of Phonon Frequencies [8-10]**

The potential V($\theta_j$) for a mode is computed corresponding to any normal coordinate $\theta_j$ of j$^{th}$ excitation, by displacing atoms in a manner specified by the eigenvector of the mode under consideration. For $|\theta_j| \leq \langle \theta_j^2 \rangle^{1/2}$ the crystal potential energy V($\theta_j$) can be written in a polynomial in $\theta_j$ as follows:

$$V(\theta_j) = a_{0,j} + a_{2,j}\theta_j^2 + a_{3,j}\theta_j^3 + a_{4,j}\theta_j^4 \qquad (S1)$$

The coefficients $a_{2,j}$, $a_{3,j}$, and $a_{4,j}$ of the harmonic and anharmonic terms in equation (S1) are useful in determining the harmonic frequencies and anharmonic shifts. The coefficients are determined by least-squares fitting of V(r$_j$) to Eq. (S1). The appropriate moment of inertia I$_j$ of the oscillator is given by

$$I_j = \sum_k m_k |u^t_{k,j}|^2 + \sum_k \underline{I_k} |u^r_{k,j}|^2$$

Here k runs over all atoms and molecules in the primitive cell and $u^t_k$ and $u^r_k$ represent the translational and rotational displacement vectors, for unit $\theta_j$, of the k$^{th}$ atom/molecule with a moment of inertia $\underline{I_k}$. The harmonic frequency is given by the expression

$$E_j = \left[\frac{2a_{2,j}}{I_j}\right]^{1/2}$$

I$_j$ and E$_j$ are the moment of inertia and calculated energy of the j$^{th}$ oscillator at 0 K. The anharmonic shifts due to third- and fourth-order terms respectively are given by the expressions

$$\Delta_j^{(3)} = -\frac{7.5\hbar}{I_j^3}\left[\frac{a_{3,j}}{E_j^2}\right]$$

$$\Delta_j^{(4)} = 3.0\hbar \left[ \frac{a_{4,j}}{(I_j E_j)^2} \right]$$

Further, the renormalized frequency $E_j(T)$ at any temperature T can be calculated as

$$E_j(T) = E_j + (\Delta_j^{(3)} + \Delta_j^{(4)})\left[ 2n_j(T) + 1 \right]$$

where $n_j(T)$ is the number of phonons excited in thermal equilibrium at any temperature given by Bose-Einstein distribution function $n_j(T)\left[ = \exp(E_j/k_B T) - 1 \right]^{-1}$.

**TABLE SI** Comparison of the experimental structural parameters [7] at 293 K in the cubic phase of $ZrW_2O_8$ with the calculations. The units used are Å for the lattice constant a and Å$^2$ for mean squared amplitudes $u^2$. The atoms are labeled as indicated in Ref. [7].

|     |       | Expt.  | Calc.  |
|-----|-------|--------|--------|
|     | a     | 9.15993| 9.3200 |
| Zr  | x     | 0.0003 | 0.0015 |
|     | $u^2$ | 0.010  | 0.012  |
| W1  | x     | 0.3412 | 0.3428 |
|     | $u^2$ | 0.012  | 0.010  |
| W2  | x     | 0.6008 | 0.6005 |
|     | $u^2$ | 0.010  | 0.008  |
| O1  | x     | 0.2071 | 0.2061 |
|     | y     | 0.4378 | 0.4387 |
|     | z     | 0.4470 | 0.4471 |
|     | $u^2$ | 0.022  | 0.020  |
| O2  | x     | 0.7876 | 0.7867 |
|     | y     | 0.5694 | 0.5675 |
|     | z     | 0.5565 | 0.5564 |
|     | $u^2$ | 0.020  | 0.018  |
| O3  | x     | 0.4916 | 0.4917 |
|     | $u^2$ | 0.023  | 0.022  |
| O4  | x     | 0.2336 | 0.2355 |
|     | $u^2$ | 0.037  | 0.034  |

**TABLE SII** Results of anharmonic calculations. The value of the parameters '$a_{3,j}$' and '$a_{4,j}$' are extracted from fitting of equation (S1) to the potential of the mode with fixed value of '$a_{2,j}$' as determined from the phonon energies.

| Representation | E (meV) | $a_{2,j}(\times 10^{-3})$ | $a_{3,j}(\times 10^{-6})$ | $a_{4,j}(\times 10^{-6})$ | $\Delta_j^{(3)}(\times 10^{-3})$ (meV) | $\Delta_j^{(4)}(\times 10^{-3})$ (meV) |
|---|---|---|---|---|---|---|
| Γ | 4.93 | 2.925 | -2.75 | 7.51 | -0.008 | 15.452 |
| Γ | 5.21 | 3.268 | -72.7 | 12.84 | -5.910 | 23.651 |
| X | 3.90 | 1.831 | 0.0 | 5.46 | 0.000 | 17.956 |
| X | 4.16 | 2.083 | 0.0 | 20.55 | 0.000 | 59.395 |
| M | 4.51 | 2.446 | 0.0 | 16.2 | 0.000 | 39.876 |
| M | 4.65 | 2.605 | 0.0 | 23.58 | 0.000 | 54.496 |
| R | 5.29 | 3.373 | 0.0 | -24.0 | 0.000 | -42.838 |

**FIG. S1** The calculated partial density of states for various atoms and the total one phonon density of states.

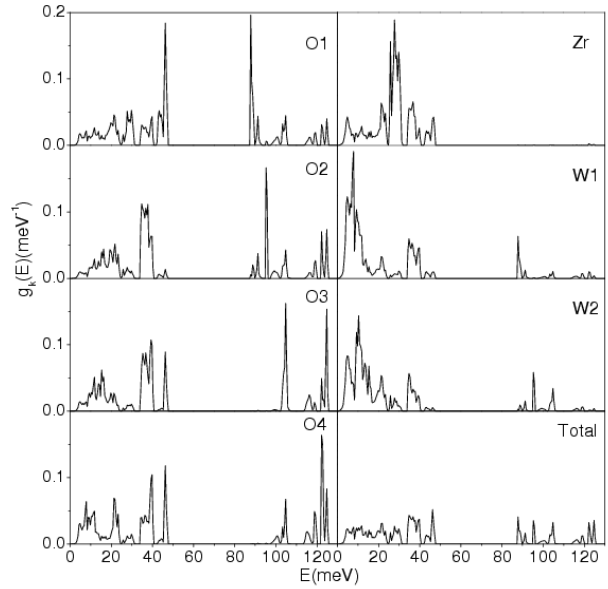

**FIG. S2** The calculated phonon dispersion curves for ZrW$_2$O$_8$ at ambient pressure. The Bradley-Cracknell notation is used for the high-symmetry points along which the dispersion relations are obtained. Γ=(0,0,0); X=(1/2,0,0); M=(1/2,1/2,0) and R=(1/2,1/2,1/2).

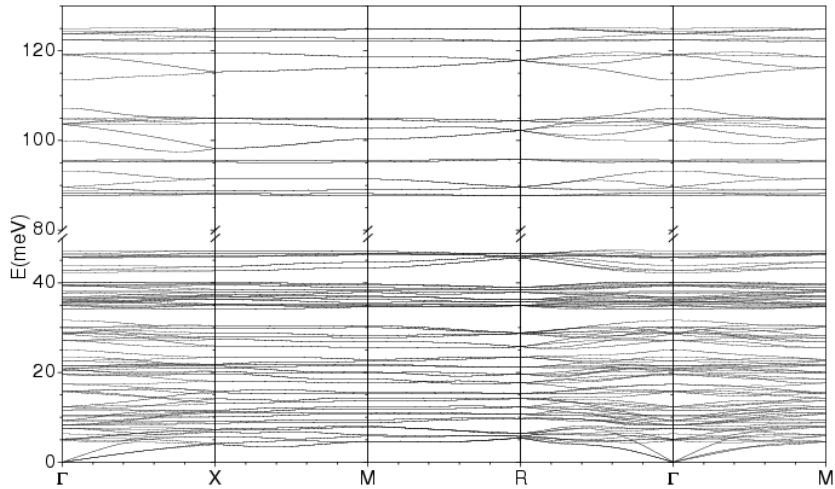

**FIG. S3** The comparison between the calculated and experimental [11,12] specific heat for the cubic $ZrW_2O_8$

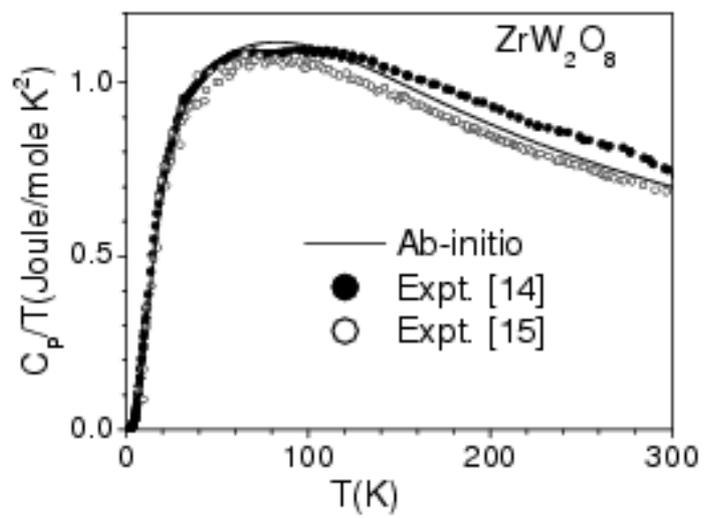

**FIG. S4** (Color online) Polarization vectors of low energy phonon modes in cubic ZrW$_2$O$_8$. Γ, X, M and R are the high symmetry points in the cubic Brillouin zone. The numbers after the mode assignments give the phonon energies and Grüneisen parameters respectively. The lengths of arrows are related to the displacements of the atoms. The atoms are labeled as indicated in Ref. [7].

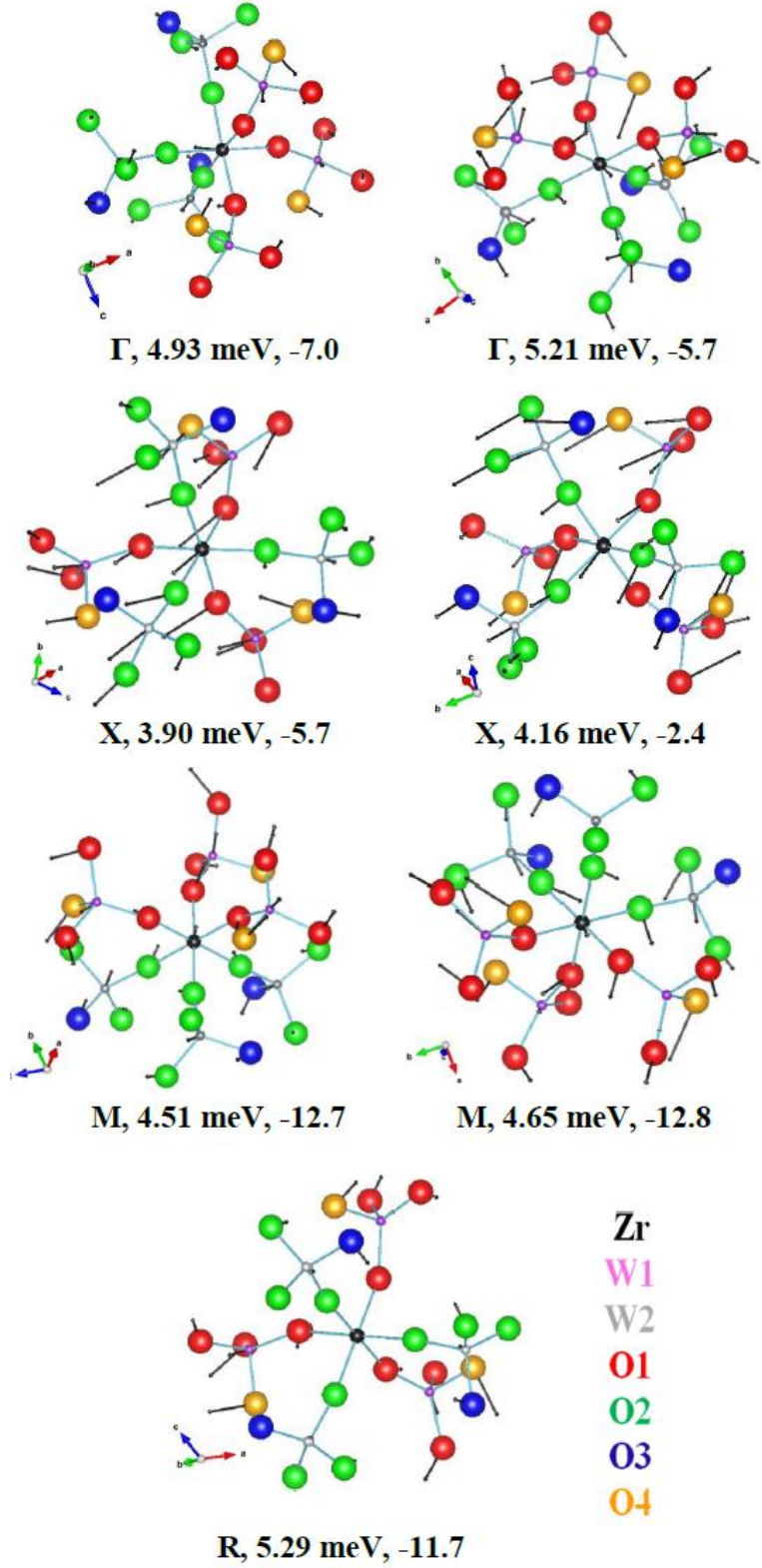

**FIG. S5** (Color online) The calculated potential wells of low energy phonon modes in cubic $ZrW_2O_8$. $\Gamma$, X, M and R are the high symmetry points in the cubic Brillouin zone. The numbers after the mode assignments give the phonon energies and Grüneisen parameters respectively.

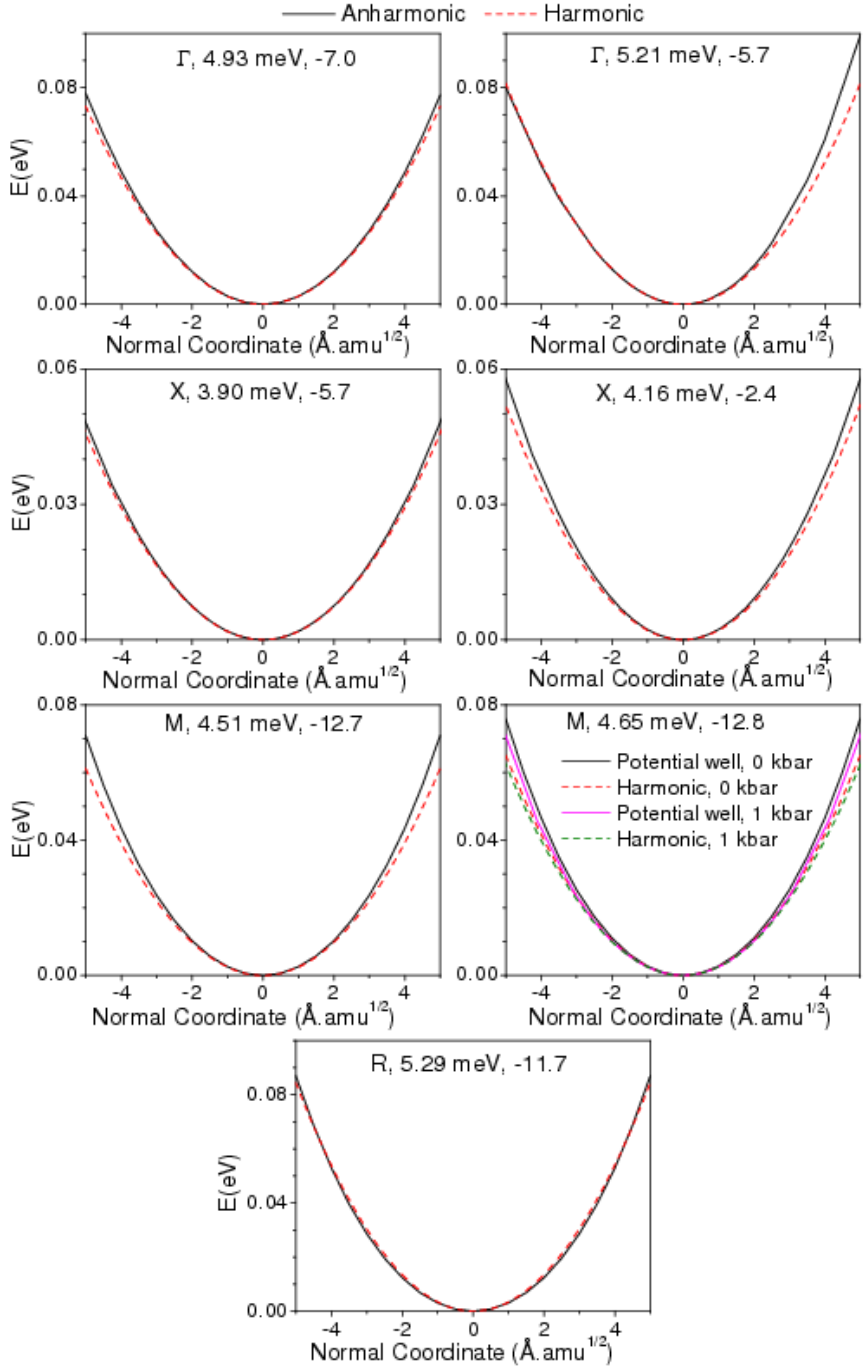

**FIG. S6** (Color online) The calculated temperature dependence of low energy phonon modes in cubic ZrW$_2$O$_8$. The temperature dependence of phonon modes is due to both the "implicit" as well as "explicit" anharmonicities. Γ, X, M and R are the high symmetry points in the cubic Brillouin zone. The numbers after the mode assignments give the phonon energies and Grüneisen parameters respectively. Solid circles are the data taken from the experimental temperature dependence of phonon density of states [13], which represents average over entire Brillouin zone. The line through the circles is guide to the eye.

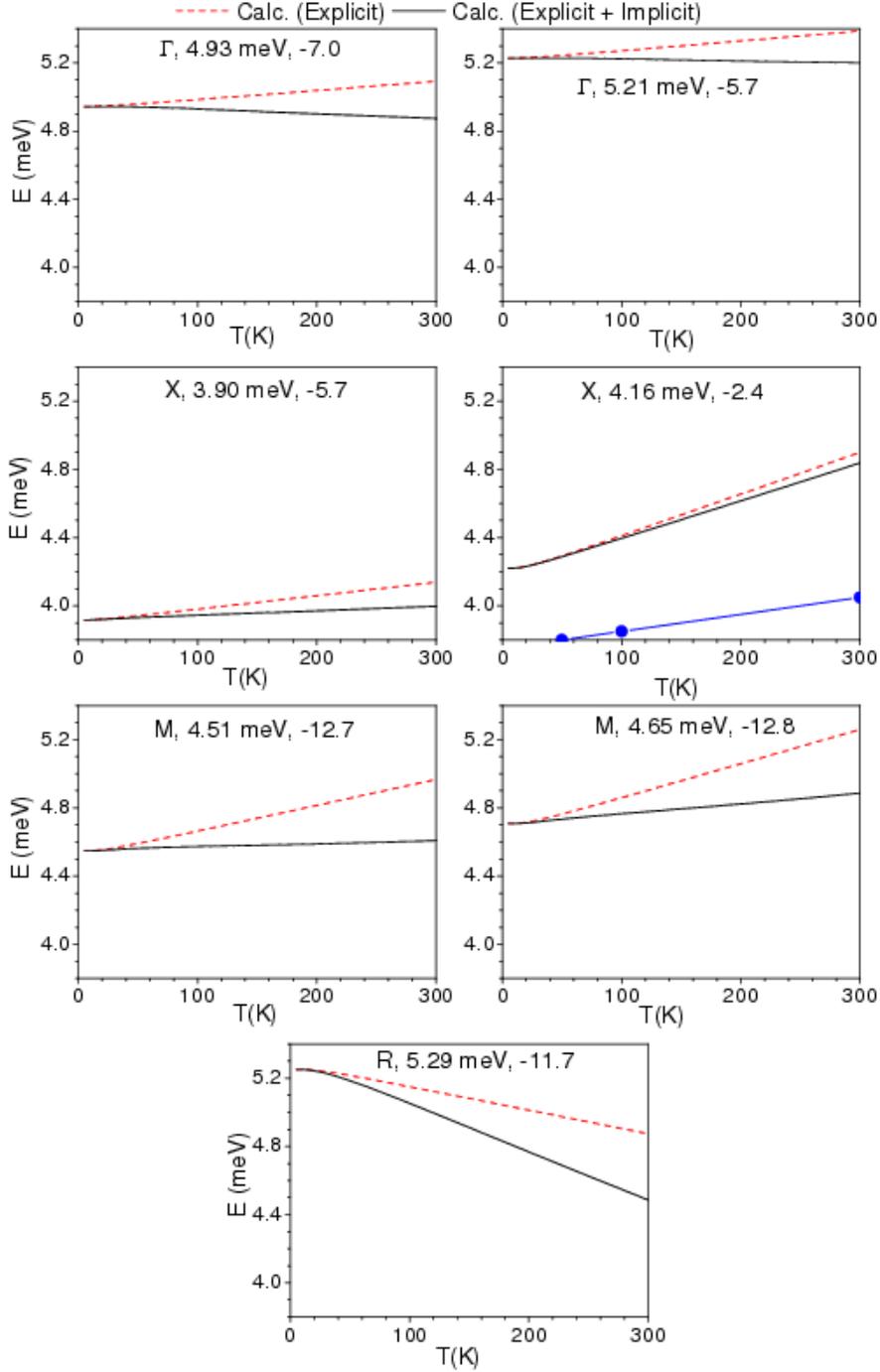